\documentclass[useAMS,usenatbib]{mn2e}
\usepackage{amsmath,fleqn,graphicx,amssymb}
\usepackage{multirow}
\renewcommand{\[}{\begin{equation}}
\renewcommand{\]}{\end{equation}}
\def\p{\partial}

\def\ex#1{\left\langle#1\right\rangle}
\def\agamaTwo{{\sc agama}b}
\def\agama{{\sc agama}}

\def\eqrf#1{(\ref{#1})}
\def\Jcrit{J_{z\rm crit}}
{\newif\ifnotend
\notendtrue
\def\veclist{ABCDEFGHIJKLMNOPQRSTUVWXYZabcdefghijklmnopqrstuvwxyz.}
\def\top#1#2.{#1}
\def\tail#1#2.{#2.}
\loop\expandafter\xdef\csname v\expandafter\top\veclist\endcsname%
{{\noexpand\bf\expandafter\top\veclist}}
\edef\veclist{\expandafter\tail\veclist}
\if\veclist.\notendfalse\fi\ifnotend\repeat}
{\newif\ifnotend
\notendtrue
\def\veclist{ABCDEFGHIJKLMNOPQRSTUVWXYZ.}
\def\top#1#2.{#1}
\def\tail#1#2.{#2.}
\loop\expandafter\xdef\csname c\expandafter\top\veclist\endcsname%
{{\noexpand\cal\expandafter\top\veclist}}
\edef\veclist{\expandafter\tail\veclist}
\if\veclist.\notendfalse\fi\ifnotend\repeat}
\def\d{{\rm d}}
\def\Vc{v_{\rm c}}
\def\cJ{{\cal J}}

\def\kpc{\,\mathrm{kpc}}

\def\kms{\,\mathrm{km\,s}^{-1}}

\def\msun{\,{\rm M}_\odot}

\def\e{\mathrm{e}}

\def\fracj#1#2{{\textstyle{#1\over#2}}}

\def\sgn{{\rm sgn}}

\title[Distribution functions for spheroids]
{Distribution functions for spheroids}

\author[]{James Binney\thanks{E-mail:
James.Binney@physics.ox.ac.uk}
\\Rudolf Peierls Centre for Theoretical Physics, Clarendon Laboratory, Oxford OX1 3PU, United Kingdom
}

\begin{document}
\maketitle

\begin{abstract}
Galaxy models comprising several components (including dark matter) that are
bound by the self-consistently generated gravitational field are readily
constructed from distribution functions (DFs) that are analytic functions of
the action integrals $\vJ$. We explain why such models have unphysical
velocity distributions unless the DFs of hot components satisfy certain
conditions as
$J_\phi\to0$. We show how DFs for both isotropic and radially biased spherical systems
can be constructed with specified $f(\vJ)$. We show how to construct DFs for
flattened systems with significant velocity anisotropy. Construction of
self-consistent models rather than populations that are confined by an
external potential leads to the conclusion that radially-biased spherical
systems are generically unstable to quadrupolar perturbations. 
Chaos is likely key to maintenance of these constraints during adiabatic
disc growth.
\end{abstract}


\section{Introduction}\label{sec:intro}

The extent and quality of the data that are available for both our Galaxy and
external galaxies has increased enormously in the last few years. In the
case of the Milky Way, the  Gaia data releases \citep{GaiaDR2general,GaiaDR3general} 
 and
data released by several spectroscopic surveys \citep{RAVEDR6,LAMOST,
APOGEE17, GalahDR4}
have enabled us to study the chemodynamics of our archetypal Galaxy in
extraordinary detail. Meanwhile, a new generation of integral field
units (IFUs) has dramatically increased our ability to dissect neighbouring
galaxies \citep[e.g.][]{CALIFA2024,MAGPI2024}. 

Notwithstanding these dramatic observational advances, the available data
remain incomplete, are marred by non-negligible uncertainties and influenced
by observational biases. The surest way to manage these limitations is to
model them with the aid of dynamical models. In its purest form this process
involves fitting a model of sufficient sophistication to the data, errors,
biases and all.

The more advanced the data are, the more sophisticated must be the models
fitted. For example, if the data suffice only to constrain the first three
moments of the distribution function (DF), namely the luminosity density, the
mean velocity and the velocity dispersion, models constructed from the Jeans
equations suffice \citep[e.g.][]{BDI,JeansMod2023,JeansMod2025}.  Gaia data yield the full velocity
distributions at large numbers of locations, and Jeans modelling cannot
adequately interpret such data. Moreover the availability of spectroscopy for
huge numbers of stars enables us to probe the dynamics of many different
populations within the Milky Way, so models need to furnish predictions for
several populations that are confined by a common gravitational potential.

Such models can be constructed in several ways. A popular approach is N-body
modelling. Unless several million particles are employed, Poisson noise in
the gravitational potential is unacceptably large \citep[e.g.][]{Aumer2016a}, so this is inevitably an
expensive technique.  Moreover, in its standard form the connection between
the initial conditions and the final model is opaque, so it is hard to fit a
model to given data.  Such fits have been achieved by the Made to Measure
(M2M) technique \citep{SyerTremaine,deLorenzi2007} in which the weights of
particles are adjusted to optimise the fit to data, but only with a great
deal of painstaking labour \citep[e.g.][]{Portail2017}.

Another popular, and slightly less labour-intensive approach to galaxy
modelling is Schwarzschild's technique \citep{Schwarzschild1979}, which has
been applied to significant numbers of galaxies in the Sauron
\citep{Emsellem2007}, {\sc califa} \citep{CALIFA}, {\sc atlas} 3D
\citep{ATLAS3D_short}, {\sc sami} \citep{SAMI}, and {\sc MANGA}
\citep{manga} surveys
\citep[e.g.][]{vdBea08,Cappellari2016,Zhu2018,CALIFA2024,MAGPI2024}. 

Here we develop a third approach to galaxy modelling, $f(\vJ)$ modelling. In
this approach, which has been applied to the Milky Way
\citep{JJB10,JJB12:dfs,Piea14_short,PifflPenoyreB,BinneyPiffl15,BinneyWong,
Binney2018,Vasiliev2019,LiBinneyRRLyrae,LiBinneyYoungD,
BinneyVasiliev2023,BinneyVasiliev2024}, to dwarf spheroidal
galaxies \citep{Pascale2018,Pascale2019,Pascale2024}, and to globular clusters
\citep{DellaCroce2024}, one assigns an analytic function of
the action integrals to each of a galaxy's components as its DF, and then
determines the observables by integrating over velocities. In early work,
the galaxy's gravitational potential was assumed up front as in Schwarzschild
modelling, but latterly the gravitational potential jointly generated by the
components has usually been obtained by solving Poisson's equation. The {\sc
agama} software package \citep{AGAMA} makes the use of the self-consistently
generated potential straightforward and affordable. 

Key for the success of $f(\vJ)$ modelling is the availability of a library of
functional forms that can be used for the DFs of disc, bulge dark and stellar
haloes, etc. Very satisfactory fits to data for the solar neighbourhood were
obtained with the `quasi-isothermal' DF for discs, which was introduced by
\cite{JJB10} and refined by \cite{JJBPJM11:dyn}. \cite{BinneyVasiliev2023}
proposed an improvement on the quasi-isothermal DF and used it to model the
large-scale structure of the Galactic disc.

Disc DFs can plausible vanish as $J_\phi\to0$ and in this way avoid a problem
with current DFs for spheroidal systems, which is the topic of this
paper. \cite{Poea15} introduced the `double-power-law' DF for spheroids, and
demonstrated that it can generate \cite{He90} and NFW \citep{NFW97} components.
\cite{Pascale2018} introduced the `exponential' form, which generates dwarf
spheroidal galaxies and used it to probe the structure of the dark haloes of
these galaxies. Unfortunately,  these forms can be safely used only for
systems with essentially isotropic velocity distributions.
Here we explain why this is the case and introduce DFs that work also in the
case of velocity anisotropy.

It is convenient to break the DF of an axisymmetric model into parts that are
even and odd functions of $J_\phi$. The odd part determines the rotation of
the model but plays no part in determining the density \citep[e.g][]{GDII}.
Here we focus on the even part, which implicitly defines a
non-rotating but possibly flattened stellar system.

Section \ref{sec:problem} defines the problem to be addressed. Section
\ref{sec:analytic} investigates it analytically.  Section \ref{sec:spherical}
solves the problem for spherical systems. Section \ref{sec:oblate} offers a
solution for flattened systems. Section \ref{sec:SelfCon} presents
self-consistent models rather than components that are confined by an
externally generated potential, finding that radial bias in a self-consistent
system flattens the even when its DF is $f(J_r,L)$.
Section~\ref{sec:adiabat} asks whether the conditions on $f$ are still
satisfied after the potential has been adiabatically deformed.
Section~\ref{sec:conclude} sums up and suggests directions for further work.

\section{The problem}\label{sec:problem}

\cite{JJB14} proposed generating anisotropic models by replacing $H(\vJ)$ as
the argument of the DF $f(H)$ of an ergodic model by a similar function of
$\vJ$ in which its components $J_i$ receive different weights: weighting an
action more heavily depopulates orbits with large values of that action
relative to the ergodic model. For example, decreasing the weighting of $J_r$
induces a radial bias in the model, while weighting $J_z$ more strongly than
$J_\phi$ flattens the model.

\cite{Poea15} showed that DFs that are double power laws in linear
combinations of the actions generate self-consistent systems that closely
mimic popular spherical system. In particular, the spherical models  introduced by
\cite{He90} and \cite{NFW97} are to a good approximation generated by a DF of the form
\[\label{eq:Posti}
f_1(\vJ)=\hbox{constant}\times
{(1+\cJ_0/\cJ)^\alpha\over(1+\cJ/\cJ_0)^\beta}\e^{-|\vJ|^2/J_{\rm cut}^2}.
\]
Here $\cJ$ is a linear combination of the actions $J_r$, $J_z$ and $J_\phi$, the exponent $\alpha$ determines the slope of the model's inner radial density
profile, $\beta$ controls its outer radial density profile, and the constant $\cJ_0$
determines the radius of the transition between them. Often $\beta$ is
too small to ensure convergence to a finite mass, so equation \eqrf{eq:Posti}
includes a Gaussian factor to truncate the model at a radius controlled by
$J_{\rm cut}$. \cite{Poea15} confined their study to spherical systems, which
are generated by making the coefficients of $J_z$ and $J_\phi$ in $\cJ$
equal, but they pointed out that,  by an extension of the
B14 proposal, flattened models can be obtained by making
the coefficient of $J_z$ larger than that of $J_\phi$.

\begin{figure}
\centerline{\includegraphics[width=.8\hsize]{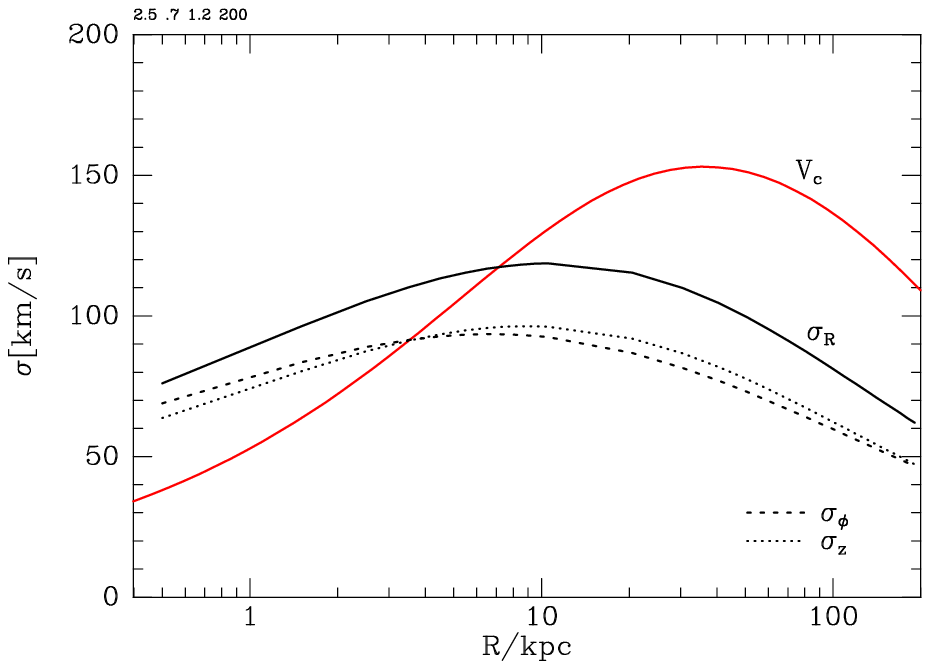}}
\centerline{\includegraphics[width=.8\hsize]{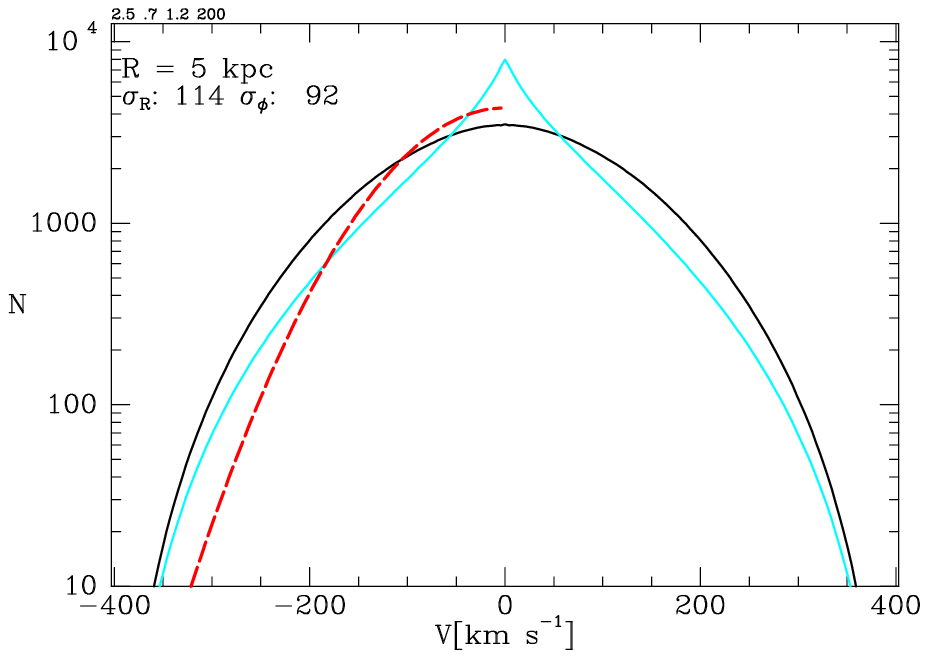}}
\caption{A halo  with a DF given by equations (\ref{eq:Posti}) and
(\ref{eq:defcJ0}) 
trapped in the potential of a similar
body flattened to axis ratio $c/a=0.5$. The lower panel shows the
distributions of $v_R$ in black and $v_\phi$ in cyan at $(R,z)=(5,0)\kpc$.
The red dashed curve shows the Gaussian with the dispersion of the $v_\phi$
distribution.}\label{fig:oldDF_squash}
\end{figure}

\begin{figure}
\centerline{\includegraphics[width=.8\hsize]{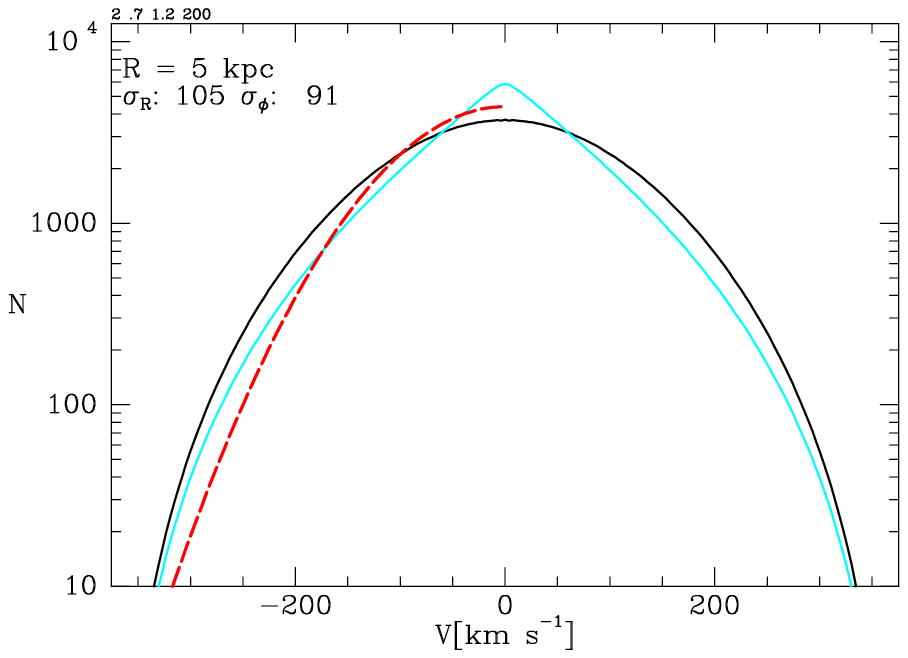}}
\caption{The distributions of radial and azimuthal velocities at $(R,z)=(5,0)\kpc$
when the same population as that shown in Fig.~\ref{fig:oldDF_squash} is
confined by the equivalent spherical potential.}\label{fig:oldDF}
\end{figure}

Fig.~\ref{fig:oldDF_squash} illustrates the problem with DFs of the type proposed by
B14. In the upper panel the red curve shows the circular speed of the
potential well in which a population is trapped. The well is generated by a
dark halo with a NFW-like radial profile and axis ratio $c/a=0.5$. The
population has a DF $f_1(\cJ)$ where
\[\label{eq:defcJ0}
\cJ=0.7J_r+1.2J_z+1.1J_\phi.
\] 
 The black curves in the upper panel show the resulting runs of the principal
velocity dispersions, $\sigma_R$ (full curve), $\sigma_z$ (dotted curve) and
$\sigma_\phi$ (dashed curve). These runs are very much what one would expect
in a system that has a modest degree of radial bias.  The lower panel shows
the distributions of $v_R$ in black and $v_\phi$ in cyan at
$(R,z)=(5,0)\kpc$. The distribution of $v_\phi$ is sharply peaked to an
extent that proves incompatible with Gaia data for the stellar halo.
Similarly peaked distributions of $v_\phi$ are found at other radii.
Fig.~\ref{fig:oldDF} shows the velocity distributions when the same
population is trapped in the equivalent spherical well. The $v_\phi$
distribution is less sharply peaked but is still more peaky than data for the
stellar halo permit (e.g., black histograms at top right of Fig.~12 in
\citealt{BinneyVasiliev2023} or in the $v_\phi$ columns of Fig.~15 in
\citealt{BinneyVasiliev2024}).

\section{Analytic considerations}\label{sec:analytic}

We now explain why some DFs yield unacceptable
distributions of $v_\phi$. 
Consider the velocity distribution $n(\vv)=f(\vx,\vv)$ at a given point $\vx$
in an ergodic model, i.e., one such that the DF can be written $f(H)$.
Then
\[
{\p f\over\p J_r}={\d f\over\d H}{\p H\over\p J_r}={\d f\over\d H}\Omega_r
\]
where $\Omega_r(\vJ)$ is the radial frequency. From the equivalent equations
for derivatives wrt $J_z$ and $J_\phi$ it follows that an ergodic DF
satisfies
\[\label{eq:OmegaRat}
{\p f/\p J_r\over\p f/\p J_\phi}={\Omega_r\over\Omega_\phi}\hbox{\quad
and\quad}
{\p f/\p J_z\over\p f/\p J_\phi}={\Omega_z\over\Omega_\phi}.
\]

The Hamiltonian $H=\fracj12 v^2+\Phi$, $f(\vv)$ is an even
function of $\vv$, so in the ergodic case $\p f/\p v_i\to0$ as $v_i\to0$. However,
from the perspective of the expression
\[\label{eq:HJ}
{\p f\over\p v_\phi}={\d f\over\d H}\left(\Omega_r{\p J_r\over\p
v_\phi}+\Omega_z{\p J_z\over\p v_\phi}+\Omega_\phi R\right),
\]
 the vanishing on $\p f/\p v_\phi$ is
puzzling because $\d f/\d H$ is generically non-zero, and the frequencies
only tend to zero as $|\vJ|\to\infty$. It follows that the last term in the
big bracket generates a non-zero value that must be cancelled by the other
two terms in the bracket. That is, as $v_\phi\to0$ at least one of $\p J_r/\p
v_\phi$ and $\p J_z/\p v_\phi$ must tend to a non-zero value that after
weighting by the frequencies magically cancels the last term.%
\citep{PifflPenoyreB}. 

\cite{PifflPenoyreB} pointed out that taking the argument of $f$ to be a
linear combination $\cJ$ of the actions as proposed by B14 prevents the three
terms in the big bracket of equation (\ref{eq:HJ}) cancelling, and thus
causes $\p f/\p v_\phi\ne0$ at $v_\phi=0$, which conflicts with both physical
intuition and Gaia data. The DF needs to be structured  such that the three
terms do cancel in the limit $v_\phi\to0$.

Satisfaction of the condition 
\[\label{eq:basic}
\lim_{v_\phi\to0}{\p f\over \p v_\phi}=0
\]
actually achieves two things. In addition to ensuring that at any location
the distribution in $v_\phi$ doesn't have a cusp or dimple at $v_\phi=0$, it
ensures that the distribution of velocity components $v_x$ and $v_y$ parallel
to the equatorial plane tends smoothly to isotropy as the symmetry axis is
approached -- this behaviour is imperative because on the axis all directions
perpendicular to the axis are equivalent. Similarly, in a spherical model
with a finite central density so the central velocity distribution is well
defined, that velocity distribution must be isotropic because all directions
are equivalent there.  

Isotropy at points on the axis is assured for any $f(\vJ)$ because there the
distributions of both $v_x$ and $v_y$ are determined by the dependence of $f$
on a single action (away from the potential's core, the relevant action is
$J_z$, while in the core it is $J_r$). As the axis is approached, the distributions
of the components $v_R$ and $v_\phi$ in the radial and azimuthal directions
should converge on the distribution of $v_x$. For this to happen, the
dependencies of $f$ on $J_z$ [which has control of $f(v_R)$] and on $J_\phi$
[which dominates $f(v_\phi)$] must converge, and ensuring that $f$ satisfies
the condition (\ref{eq:basic}) also imposes this convergence. 

\subsection{Spherical models}\label{sec:anal_sph}

Let $v_{\rm t}$ be the magnitude of the  tangential component of velocity in
a spherical system, and ask how $\p f/\p v_{\rm t}$ contrives to vanish as
$v_{\rm t}\to0$ in a spherical model.  Any spherical model has a DF
that can in principle be expressed as $f(J_r,L)$, where the total angular
momentum is
\[
L=rv_{\rm t}=J_z+|J_\phi|.
\]
In the ergodic case, equation (\ref{eq:HJ}) can
be written
\begin{align}\label{eq:HJ2}
{\p f\over\p v_{\rm t}}&={\d f\over\d H}\left(\Omega_r{\p J_r\over\p
v_{\rm t}}+\Omega_{\rm t}{\p L\over\p v_{\rm t}}\right)\cr
&={\d f\over\d H}\left(\Omega_r{\p J_r\over\p
v_{\rm t}}+\Omega_{\rm t}r\right).
\end{align}
Differentiating the defining equation  of $J_r$, we have
\[\label{eq:pJrpvphi}
{\p J_r\over\p v_{\rm t}}\bigg|_r
={v_{\rm t}\over\pi}\int_{r_{\rm p}}^{r_{\rm a}}\d r'\,
{1-r^2/r^{\prime2}\over\sqrt{2(E-\Phi)-L^2/r^{\prime2}}},
\]
 where $r_{\rm p}\le r_{\rm a}$ are the roots of the bottom of the integrand.
As $v_{\rm t}\to0$,
$r_{\rm p}$ tends to zero and the contribution to the
integral from the second term on the top becomes large for any finite $r$. In
fact, the product of the rising integral and the prefactor $v_{\rm t}$  tends to a
non-zero limit.  Appendix \ref{app:limit} shows that
\[\label{eq:dJrdv}
{\p J_r\over\p v_{\rm t}}\bigg|_r
=-\fracj12r
\]
so the two terms in the bracket of
equation (\ref{eq:HJ2}) do cancel provided $\Omega_r=2\Omega_{\rm t}$.

In the general spherical case we have
\[
{\p f\over\p v_{\rm t}}={\p f\over\p J_r}{\p J_r\over\p v_{\rm t}}+{\p f\over\p
L}r.
\]
Since $\p J_r/\p v_{\rm t}$ is given by equation (\ref{eq:dJrdv}), the
terms on the right will cancel as $v_{\rm t}\to0$ iff
\[\label{eq:sphLim}
\lim_{L\to0}{\p f/\p J_r\over\p f/\p L}=2=\lim_{L\to0}{\Omega_{\rm t}\over\Omega_r}.
\]
 Since the ratio on the right is precisely that appearing in the ergodic
relations (\ref{eq:OmegaRat}), we conclude that any physically acceptable DF of the form $f(J_r,L)$ must tend to the
ergodic DF as $L\to0$

This limiting condition achieves two things. In addition to its stated achievement
it ensures that as the centre is approached, the velocity distribution in any
spherical model tends smoothly to the isotropic central distribution. The
latter depends only on the dependence of $f$ on $J_r$ at $L=0$, while
infinitesimally away from the centre the dependence of $f$ on $L$ has a big
influence on the distribution of tangential components $v_{\rm t}$. A smooth
transition to isotropy at the centre is possible only if the
structure of the DF for small $L$ is constrained by equation \eqrf{eq:sphLim}.

\subsection{Flattened models}\label{sec:anal_flat}

Consider now the kinematics of a model with DF $f(\vJ)$ that is confined by a
possibly flattened potential. Instead of equation (\ref{eq:HJ}) we now have
\[\label{eq:fJ}
{\p f\over\p v_\phi}={\p f\over\p J_r}{\p J_r\over\p v_\phi}+
{\p f\over\p J_z}{\p J_z\over\p v_\phi}+{\p f\over\p J_\phi}R,
\]
where $(R,z,\phi)$ are cylindrical polar coordinates.
Consideration of how this equation works as $v_\phi\to0$ is helped if we
return to our analysis of the spherical case with $f(J_r,L)$ but computing $\p f/\p
v_\phi$ rather than $\p f/\p v_{\rm t}$. One easily shows that
\begin{itemize}
\item  $\p L/\p v_\phi\to0$ as $v_\phi\to0$ at $J_z>0$ but $\p L/\p
v_\phi\to\pm R$ when $J_z=0$;
\item the derivative of $L$ behaves thus because in equation (\ref{eq:fJ})
the second and third terms on the right usually cancel because
\[
{\p J_z\over\p v_\phi}={\p (L-|J_\phi|)\over\p v_\phi}={r^2 v_\phi\over
L}\mp R.
\]
\item $\p J_r/\p v_\phi\to0$  as $v_\phi\to\pm0$ at $J_z>0$ but $\p J_r/\p
v_\phi\to\mp R/2$ when $J_z=0$.
\end{itemize}
These results suggest that in a flattened potential two different
cancellations are required to ensure that $\p f/\p v_\phi$ vanishes with
$v_\phi$. In a spherical model the third term on the right of equation
\eqrf{eq:fJ} is cancelled by the first term
when $J_z=0$, while in a flattened potential this cancellation occurs for
$J_z$ less than a critical value $\Jcrit$. In a spherical potential,
the second and third terms on the right of equation
(\ref{eq:fJ}) cancel for $J_z>0$ because $\p f/\p J_z=\p f/\p J_\phi$ and $\p L/\p
v_\phi\to0$ when $J_z>0$, while in a flattened potential this cancellation
occurs for $J_z>\Jcrit$ only.

\begin{figure}
\centerline{\includegraphics[height=.48\hsize]{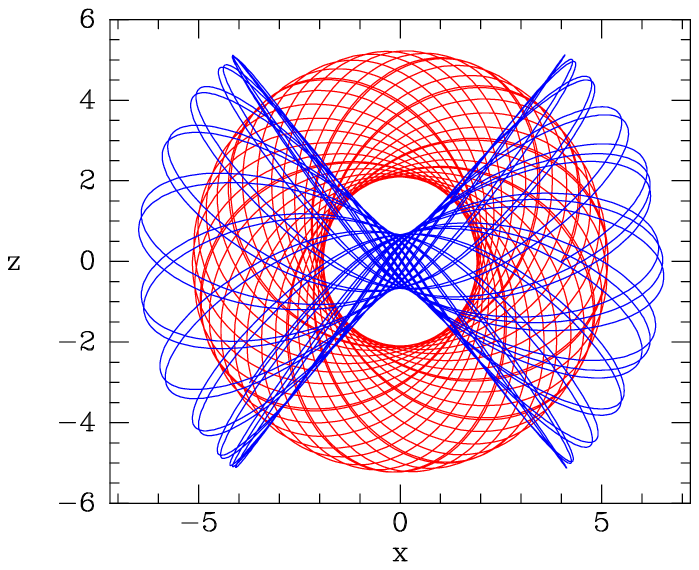}\quad
\includegraphics[height=.48\hsize]{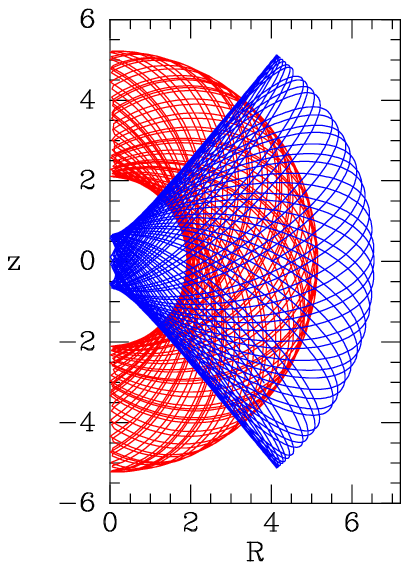}}
\caption{Left panel: two orbits of the same energy in a perfect ellipsoid at $L_z=0$. The red (loop) orbit
has $J_z>\Jcrit$ while the blue (box) orbit has $J_z<\Jcrit$. Right panel: two orbits that differ from those in the left panel in
having small but non-zero $L_z$ and hence must be in the meridional plane.}\label{fig:BoxLoop}
\end{figure}

To understand this situation, it helps to consider orbits in a flattened
potential with vanishing $J_\phi$ \citep{WrightB}. These are conveniently
studied in the $(x,z)$ plane (Fig.~\ref{fig:BoxLoop}), $y$ being always zero.
In this plane a flattened potential is elongated along the $x$ axis and $J_z$
plays the role of an generalised angular momentum. There is a critical value
$\Jcrit$ of $J_z$ such that orbits with $J_z>\Jcrit$ circulate,
while those with $J_z<\Jcrit$ librate about the $x$ axis: one speaks of
loop and box orbits. Fig.~\ref{fig:BoxLoop} illustrates this situation by
plotting box orbits in blue and loop orbits in red both at vanishing $J_\phi$
(left panel) and at non-zero $J_\phi$ (right panel).

When $J_\phi=0$, the azimuthal coordinate switches from $\phi=0$
when $x>0$ to $\phi=\pi$ when $x<0$. It is not hard to see  that on a
loop orbit the period of these oscillations in $\phi$ coincides with the
period of the oscillations in both $x$ and $z$, while on a box orbit the
$\phi$ period equals the period in $x$ but is typically longer than the
period in $z$. That is, when $J_\phi=0$
\[\label{eq:loopBox}
\Omega_\phi=\begin{cases}
\Omega_x=\Omega_z&\hbox{when }J_z>\Jcrit\hbox{ (loop orbits)}\cr
\Omega_x< \Omega_z&\hbox{when }J_z<\Jcrit\hbox{ (box orbits)}
\end{cases}
\]
Since $R=|x|$,  on box orbits $\Omega_r=2\Omega_x$ and $\Omega_\phi=\Omega_r/2$.

Consider now orbits with non-zero $J_\phi$. When $J_\phi\ne0$ an orbit cannot
reach the symmetry axis and it is no longer confined to a plane. When $J_\phi$
is small we have very anharmonic evolution in $\phi$ in that it changes
slowly until the star approaches the axis, and then $\phi$ rapidly increases by
$\sim\pi$ as the star passes the axis before evolving slowly as the star moves out
to apocentre and back. Since the frequencies must be continuous functions on
action space, $\Omega_\phi$ must lie close to $\Omega_r/2$ if $J_z$ is small,
and close to $\Omega_z$ for larger $J_z$. That is, the relationship
\eqrf{eq:loopBox} must be approximately valid even for non-zero $J_\phi$ with
$\Jcrit$ now identified from the run of orbital frequencies.

Applying these limiting forms of the frequency ratios to the ergodic
relations (\ref{eq:OmegaRat}), we find that an ergodic DF must be such that
\begin{align}\label{eq:keyConds}
{\p f/\p J_z\over\p f/\p J_\phi}&\to1\hbox{ as }J_\phi\to0\hbox{ with }J_z>\Jcrit\cr
{\p f/\p J_r\over\p f/\p J_\phi}&\to2\hbox{ as }J_\phi\to0\hbox{ with }J_z<\Jcrit.
\end{align}
 For these limiting ratios of derivatives in equation 
to lead to required cancellations in equation \eqrf{eq:fJ}, we must have
that
\[\label{eq:pJzpVphi}
\lim_{v_\phi\to+0}{\p J_z\over\p v_\phi}
=-\lim_{v_\phi\to+0}{\p J_\phi\over\p v_\phi}\quad (J_z>\Jcrit)
\]
and
\[\label{eq:pJrpVphi}
\lim_{v_\phi\to+0}{\p J_r\over\p v_\phi}
=-\fracj12\lim_{v_\phi\to+0}{\p J_\phi\over\p v_\phi}\quad (J_z<\Jcrit).
\]
By considering an ergodic model we have derived relationships,
\eqrf{eq:pJzpVphi} and \eqrf{eq:pJrpVphi}, that are set by the potential
alone and can be applied to models with any DF. 

Given that the derivatives of $\vJ$ that appear in equation \eqrf{eq:fJ}
satisfy equations \eqrf{eq:pJzpVphi} and \eqrf{eq:pJrpVphi}, it is now clear
that the DF of even an anisotropic model must satisfy
\[\label{eq:condF1}
\lim_{J_\phi\to0}{\p f/\p J_z\over\p f/\p J_\phi}=1
\quad (J_z>\Jcrit).
\] 
and
\[\label{eq:condF2}
\lim_{J_\phi\to0}{\p f/\p J_r\over\p f/\p
J_\phi}=2\quad (J_z<\Jcrit)
\]

\subsection{Summary}

At the beginning of this section we saw that $\p f/\p v_\phi$ will not vanish
with $v_\phi$ unless certain products of derivatives of $f$ and of the actions
cancel nicely. We further saw that for such cancellation to occur the partial
derivatives of $f$ as $J_\phi\to0$ must satisfy $\p f/\p J_r=\fracj12\p f/\p
J_\phi$ at $J_z$ less than a critical value and $\p f/\p J_z=\p f/\p J_\phi$
at lager $J_z$. These conditions should ensure that $\p f/\p v_\phi$ vanishes
by virtue of a relation that we inferred between derivatives of the actions
with respect to $v_\phi$.

\begin{figure}
\centerline{\includegraphics[width=\hsize]{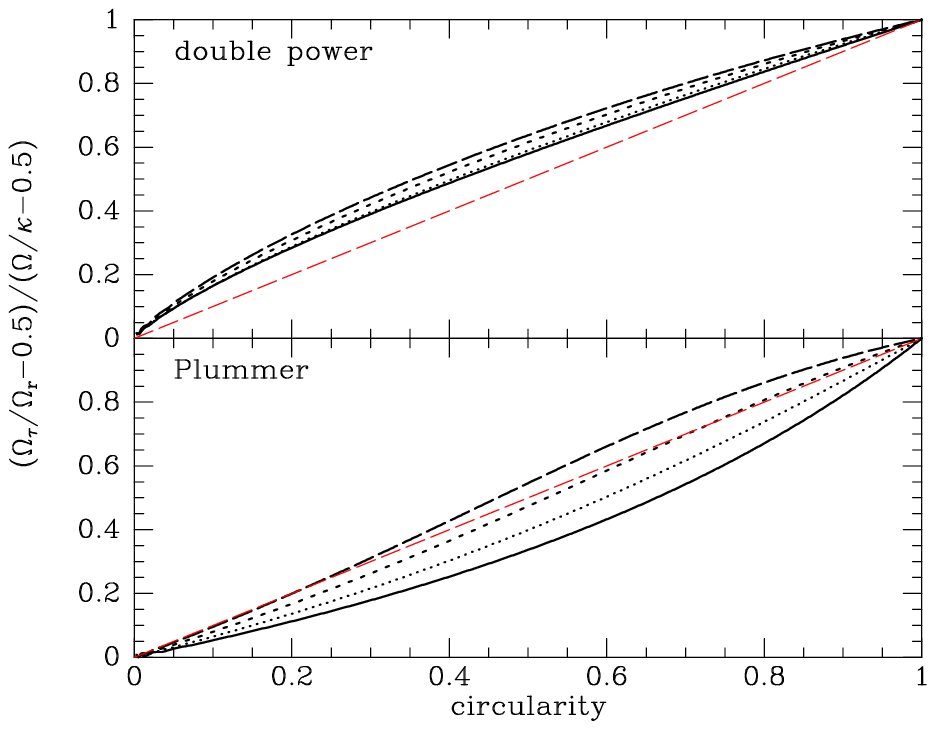}}
\caption{Each black curve shows how $\Omega_\phi/\Omega_r$ varies with
circularity $c$ at a fixed energy in a dark-halo like potential (upper panel)
or a Plummer potential (lower panel). The red lines show
$y=x$.}\label{fig:ham1}
\end{figure}
\section{DFs for spherical models}\label{sec:spherical}

We now give an algorithm for defining DFs $f(\vJ)$ that produce acceptable
spherical models. Given a point $(J_r,L)$ in reduced action space, we
integrate the ODE
\[\label{eq:basicDE}
{\d L\over\d J_r}=-g(\vJ),
\]
where $g>0$ is some positive function, in the direction of decreasing $J_r$
and increasing $L$ until $J_r=0$ and $L=L_c$.  Then we set the value
$f(J_r,L)$ of the DF at our original point to the value $f_0(L_c)$ taken by a
non-negative monotone decreasing function $f_0$ of one variable. In the
case of a dark halo, a suitable
choice for $f_0$ would be
\[\label{eq:Posti1d}
f_0(L)=\hbox{constant}\times{(1+J_0/L)^\alpha\over(1+L/J_0)^\beta}\e^{-L^2/J_{\rm
cut}^2},
\]
 which is the DF defined by equation (\ref{eq:Posti}) evaluated on
$\vJ=(0,0,L)$. If the target were a dwarf spheroidal the exponential DF introduced by
\cite{Pascale2018} would be our starting point but the procedure would be the
same. 

With this algorithm, the DF is defined by the functions $g$ and $f_0$. The
path of integration set by $g$ defines a series of orbits of decreasing
eccentricity that all have the same phase-space density. The function $f_0$
says what this density is.

If we set $g=\Omega_r/\Omega_{\rm t}$, the Hamiltonian will be constant along
the integration path, and the DF will be that of an ergodic model. If we set
$g<\Omega_r/\Omega_{\rm t}$, $L$ will increase more slowly as  $J_r$
decreases than in the ergodic case, with the consequence that our final value
$L_c$ of $L$ will be smaller than in the ergodic case and the DF
$f(J_r,L)=f_0(L_c)$ will be larger than in the ergodic model. Thus setting $g<\Omega_r/\Omega_{\rm
t}$ generates a radially biased model, while setting $g>\Omega_r/\Omega_{\rm
t}$ generates a tangentially biased model.

Regardless of what type of model we are generating, $g$ must satisfy
$\lim_{L\to0}g=2$ so $\p f/\p v_{\rm t}$ vanishes with $v_{\rm t}$.

\subsection{Ergodic models}

We now use this algorithm to construct an ergodic model of a dark halo.  This
goal is achieved by setting $g$ to an analytic approximation to
$\Omega_r/\Omega_{\rm t}$.
We use as our measure of an orbit's circularity 
\[
c\equiv{L\over L+J_r},
\]
which ranges from zero for radial orbits to unity for circular orbits.  Each
curve in Fig.~\ref{fig:ham1} shows $\Omega_{\rm t}/\Omega_r$ as a function of
circularity along a sequence of orbits at a given energy. The curves in the
lower panel are for orbits confined by a Plummer potential, while those in
the upper panel are for orbits confined by a double-power-law potential,
similar to that of a baryon-free dark halo. These two potentials differ from
one another as much as any two galactic potentials, yet in each case
$\Omega_{\rm t}/\Omega_r$ is not far from a linear function of the circularity
$c$, so we have
\[\label{eq:approxOmegas}
{\Omega_{\rm
t}\over\Omega_r}\simeq\fracj12+c\Big({\Omega\over\kappa}-\fracj12\Big),
\]
 where $\Omega$ and $\kappa$ are the epicycle frequencies.
In the case of the double-power potential, we could refine this approximation
by fitting a quadratic or cubic in $c$ to the black curves in the upper panel
of Fig.~\ref{fig:ham1}, but the examples below indicate that this is
unnecessary.

The ratio of epicycle frequencies appearing in equation
(\ref{eq:approxOmegas}) is evaluated at the energy of the sequence of orbits
through which we are integrating. From the potential we can obtain a
expression for the ratio in terms of $L_c(E)$, but $L=L_c$ is the end-point of
the integration. Hence we are obliged to guess its value. A sensible guess is
$L_c\simeq L+1.4J_r$. Fortunately, the DF obtained is extremely insensitive
to this guess because the ratio of epicycle frequencies (in contrast to the
frequencies themselves) is a weak function of
$L_c$.

\begin{figure*}
\centerline{
\includegraphics[width=.32\hsize]{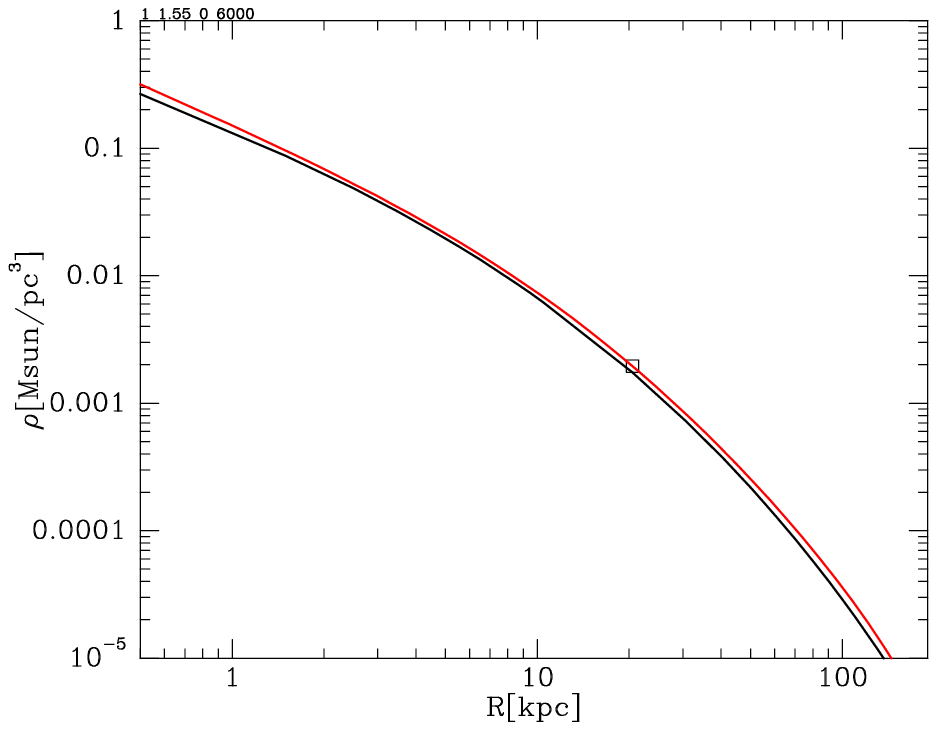}
\includegraphics[width=.32\hsize]{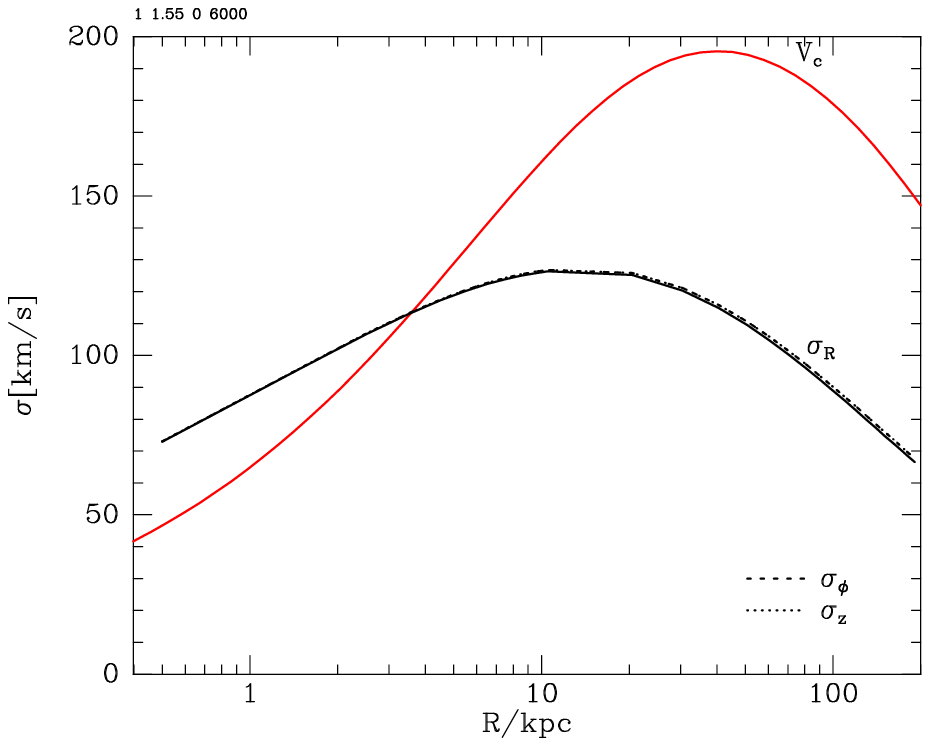}
\includegraphics[width=.32\hsize]{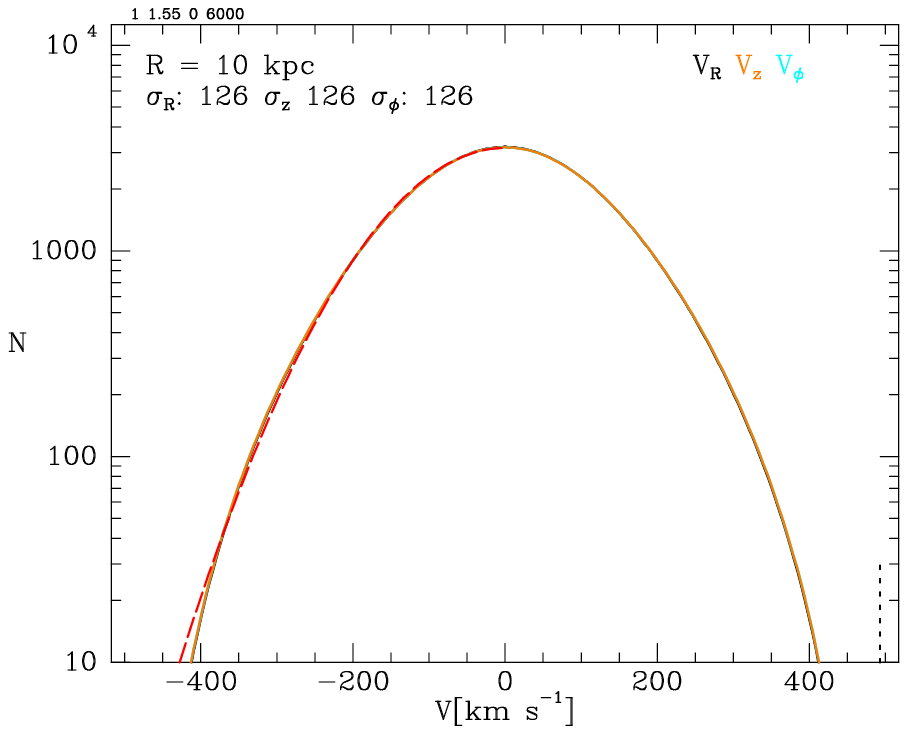}}
\centerline{
\includegraphics[width=.32\hsize]{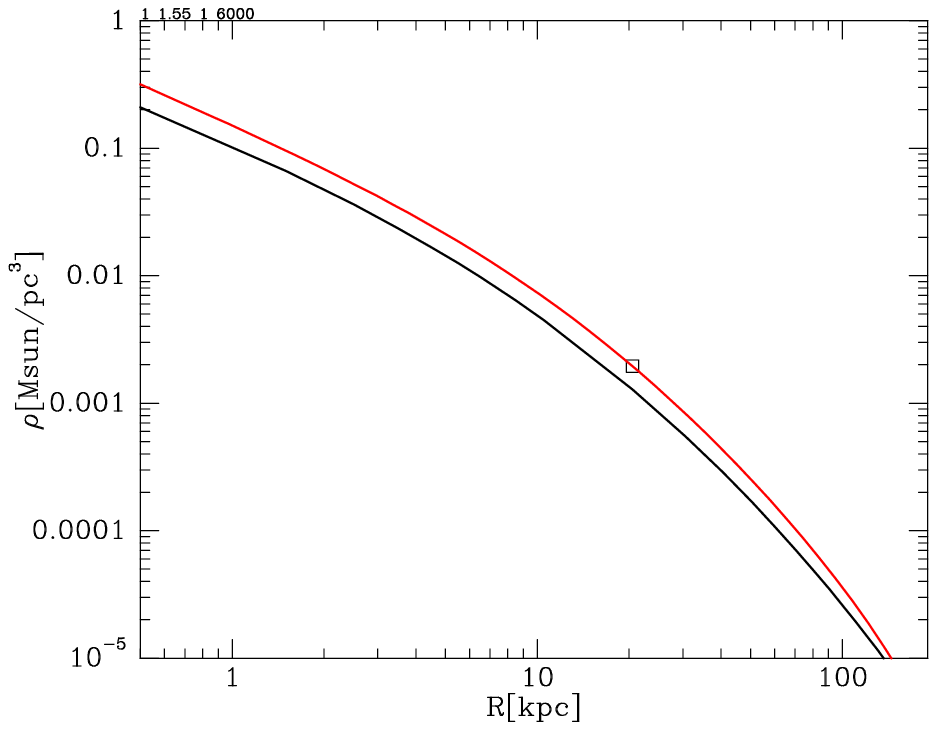}
\includegraphics[width=.32\hsize]{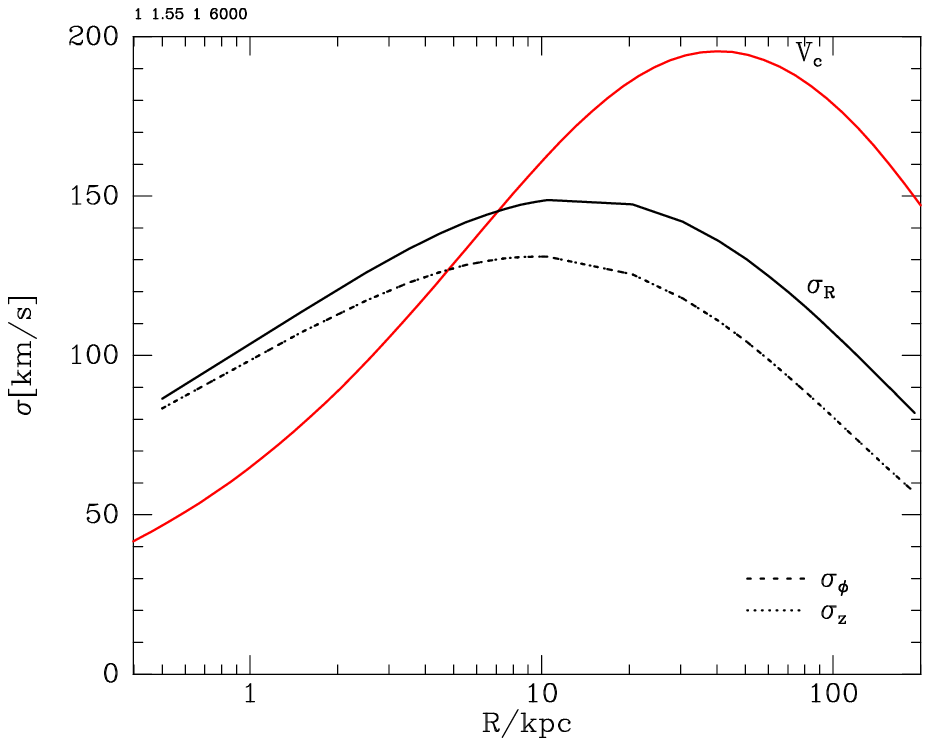}
\includegraphics[width=.32\hsize]{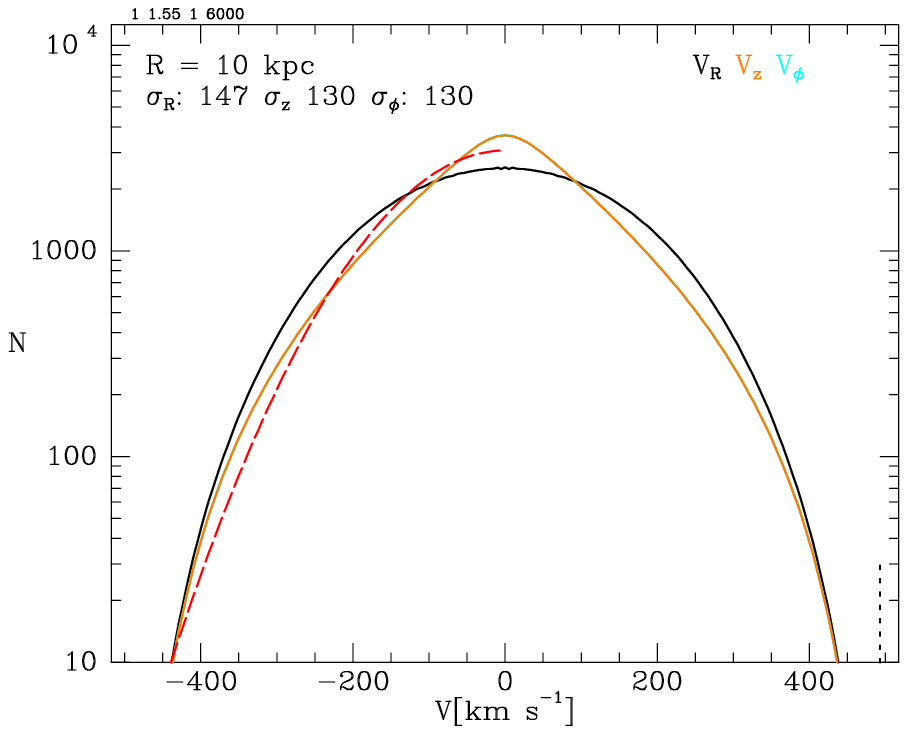}}
\centerline{
\includegraphics[width=.32\hsize]{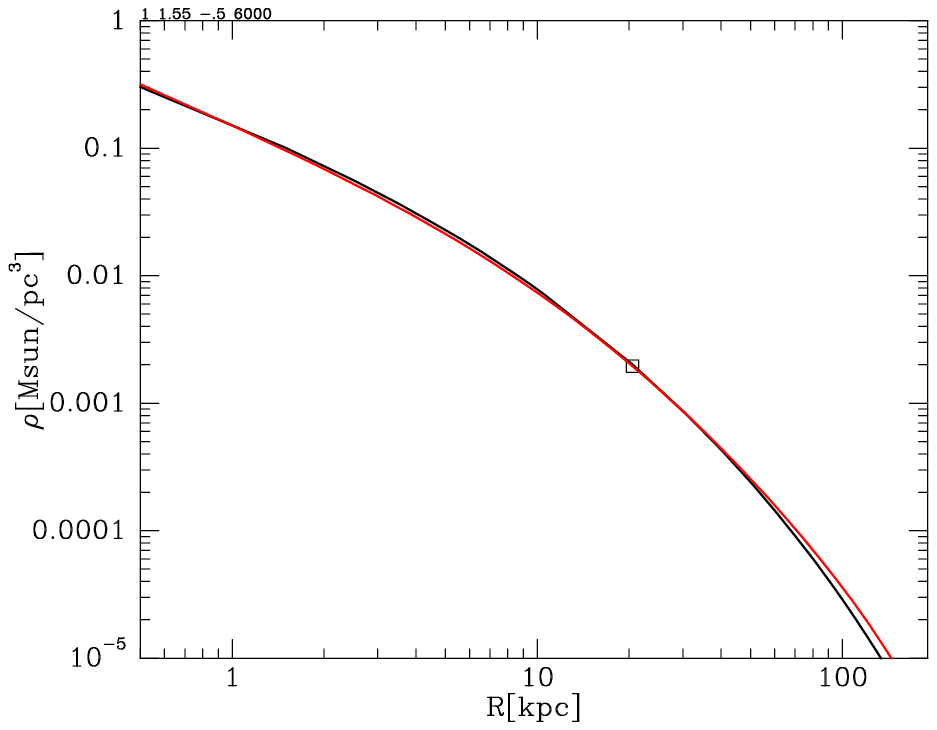}
\includegraphics[width=.32\hsize]{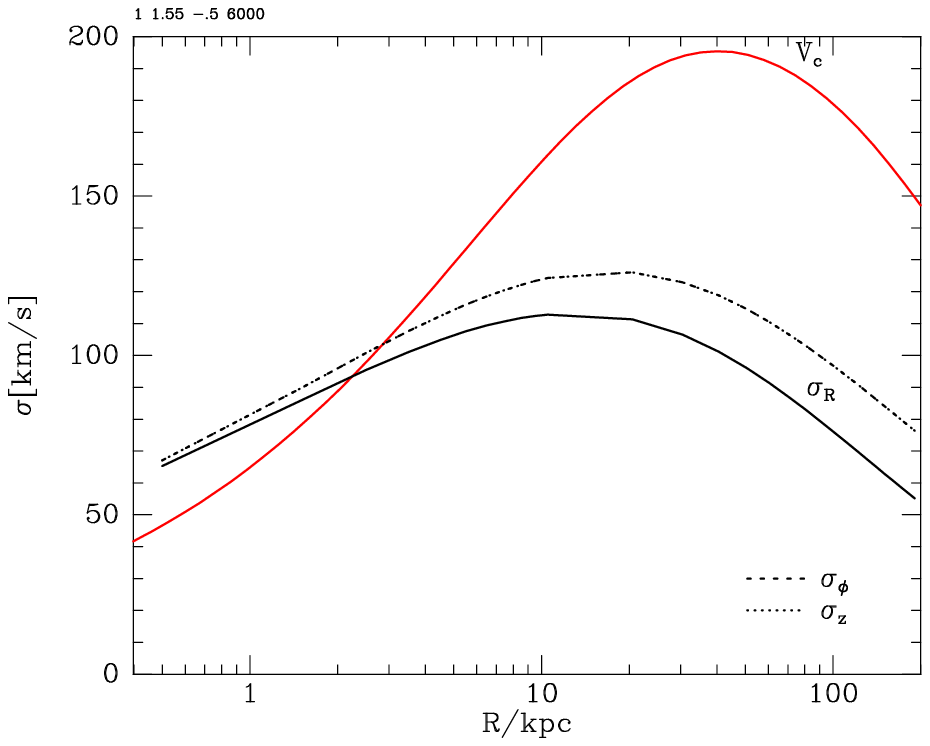}
\includegraphics[width=.32\hsize]{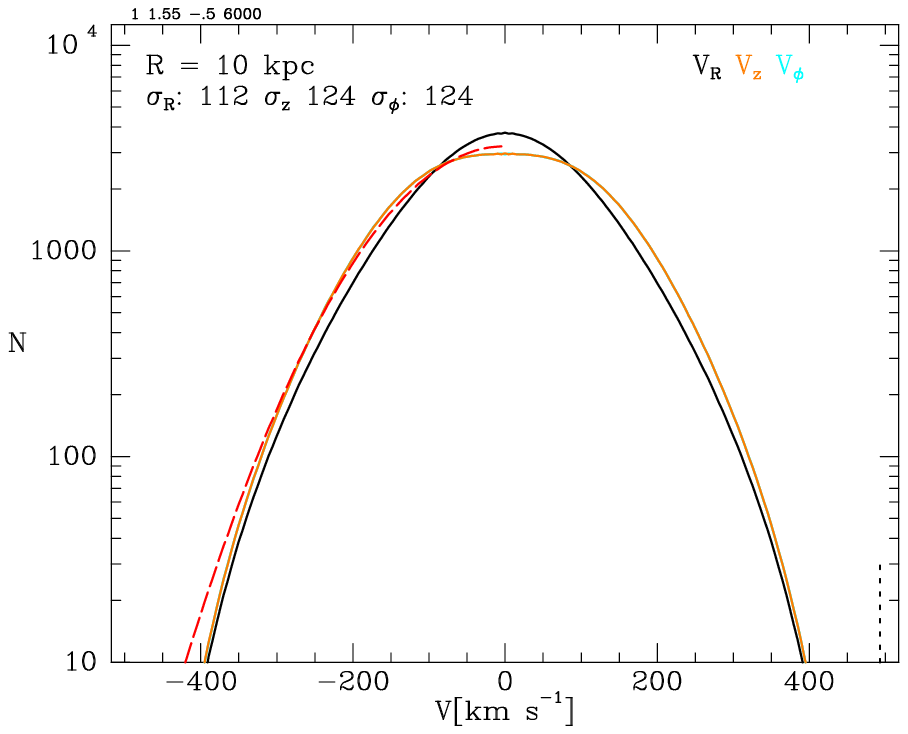}}
 \caption{Top row: a population of test particles with DF $f(h_0[\vJ])$
trapped in a spherical double power-law potential. In the left panel the red
curve shows the density distribution that generates the potential and the
black curve shows the density of test particles. The square marks the point
at which $\d\ln\rho/\d\ln r=-2$. The middle panel shows the radial and
tangential velocity dispersions in this isotropic model. In the right
panel black and overplotted orange  curves show the distribution of $v_R$
and $v_{\rm t}$
velocity components $10\kpc$ from the centre.  The dashed red curve shows the
Gaussian with the same standard deviation as $N(v_{\rm t})$.  Centre row: the
same diagnostics for a population of test particles with DF $f(h_2[\vJ])$
with $\beta_v=1$ trapped in the same potential. In the right panel the dashed
red curve shows the Gaussian with the same standard deviation as he orange
distribution of $v_{\rm t}$. Bottom row: as the middle row but for a
tangentially biased population
with $f(h_1[\vJ])$ with negative $\beta_v=-0.5$.}\label{fig:iso}
\end{figure*}

\begin{figure}
\centerline{\includegraphics[width=.8\hsize]{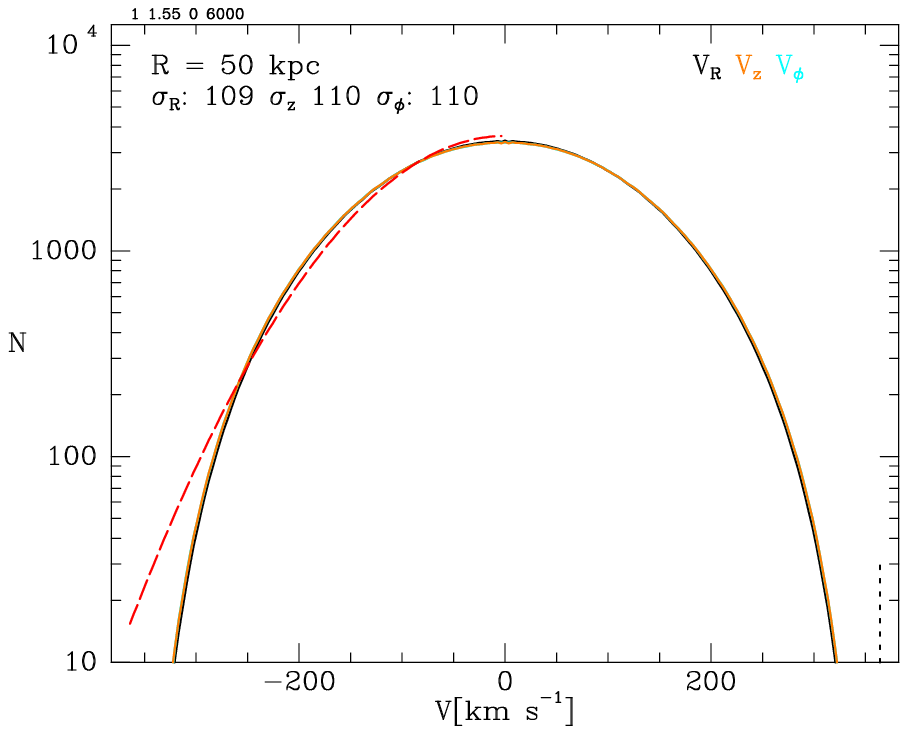}}
\centerline{\includegraphics[width=.8\hsize]{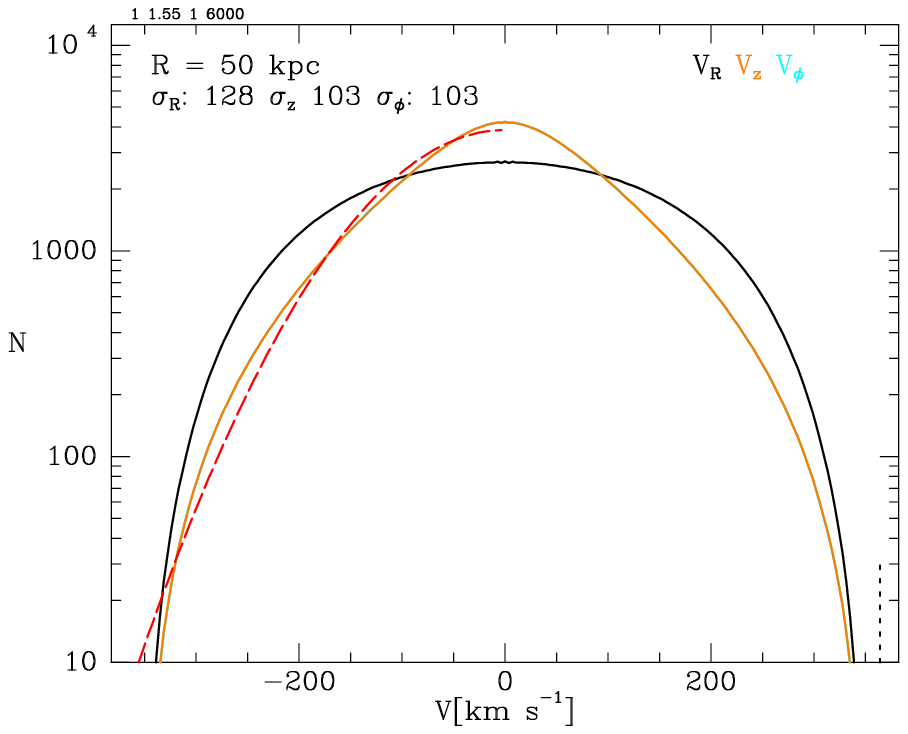}}
\caption{The same as the top right and middle right  panels of Fig.~\ref{fig:iso} but at
$r=50\kpc$. In the ergodic model the distribution of $v_z$, plotted in
orange, almost completely
obscures those of $v_R$ and $v_\phi$, plotted in black and cyan. The distributions differ
significantly from the Gaussian with the same standard deviation, plotted in
broken red. In the radially biased model plotted below, the $v_{\rm t}$ profile is pointed
while that of $v_R$ is very flat-topped.}\label{fig:iso2}
\end{figure}

The top row of Fig.~\ref{fig:iso}  shows the
structure of a population of test particles that have the DF defined by
\[\label{eq:iso_g}
g(\vJ)=\bigg[\fracj12+c\Big({\Omega\over\kappa}-\fracj12\Big)\bigg]^{-1},
\]
 with the constant $\Omega/\kappa$ evaluated at $L_c=L+1.4J_r$ and $f_0$ given by equation
(\ref{eq:Posti1d}) when they are trapped in a spherical double-power-law
potential scaled to resemble the potential of our Galaxy's dark halo.
Table~\ref{tab:DFs} gives details of the DF. The distributions of $v_z$ and
$v_\phi$ are identical and not distinguishable from the distribution of
$v_R$. In fact, in the top right panel the distribution of $v_z$ at
$r=10\kpc$, plotted in orange, obscures the distributions of $v_R$ and $v_\phi$,
plotted in black and cyan, respectively. The distributions are remarkably
Gaussian: the dashed red curve is the Gaussian with the same standard
deviation, $\sigma_R=126\kms$, as the actual $v_\phi$ distribution. The velocity
distributions at $R=50\kpc$ plotted in Fig.~\ref{fig:iso2} deviate more from
a Gaussian because the truncation of the $v$ distribution at $v_{\rm
esc}=\sqrt{-2\Phi}$ is more stringent -- the dashed vertical line at right
bottom of the velocity histograms marks $v_{\rm esc}$.

\begin{table}
\caption{Parameters of the DFs of Fig.~\ref{fig:iso}.}\label{tab:DFs}
\begin{tabular}{ccccc}
$J_0$ & $J_{\rm cut}$ & $\alpha$ & $\beta$ & $\beta_v$\\
\hline
6000&25000&1.55&2.45&0\\
6000&25000&1.55&2.45&1\\
6000&25000&1.55&2.45&$-0.5$\\
\end{tabular}
\end{table}

\subsection{Introducing velocity anisotropy}\label{sec:sph_aniso}

As explained above, to generate a radially-biased model we should take
$g<\Omega_r/\Omega_{\rm t}$, while continuing to satisfy the constraint
$\lim_{L\to0}g=2$. A possible choice of $g$ is
\[\label{eq:anisoDE}
g_\beta(J_r,L)=g_H(J_r,L)\exp[-\beta_v\sin(c\pi/2)],
\]
where $g_H$ is given by equation (\ref{eq:iso_g}).  The
factor $\exp[-\beta_v\sin(c\pi/2)]$ multiplying $g_H$ causes $g$ to be less
than $g_H$ when $\beta_v>0$ and greater than $g_H$ when $\beta_v<0$, and thus creates
radial/tangential bias depending on the sign of $\beta_v$.  The sine within the
exponential ensures satisfaction of the fundamental requirement that $g\to
g_H$ as $L\to0$. The choice of a sine is arbitrary: any
function that vanishes with its argument and has a finite range might be
used.

The centre row of panels in Fig.~\ref{fig:iso} shows the structure of a
population with this DF when $\beta_v=1$. The left panel of the centre row
shows that moving $\beta_v$ to a positive value has only a minor impact on the
population's radial density profile but the centre panel shows that it causes
$\sigma_R>\sigma_{\rm t}$. The right panel shows that the difference,
$147/130\kms$, in the dispersions at $R=10\kpc$ under-plays the difference
between the $v_R$ and $v_{\rm t}$ velocity distributions. Now both
distributions deviate significantly from Gaussians: the shift of orbits
towards higher eccentricity has depressed the orange curve for $N(v_{\rm t})$
around $v\simeq v_{\rm c}$ and elevated it around $v_{\rm t}\sim0$ by
replacing circular orbits with eccentric orbits seen near apocentre. Note that
at large $|v|$ the $v_R$ and $v_{\rm t}$ distributions converge. This is
because at high speeds eccentric orbits seen at pericentre boost $N(v_{\rm
t})$. The lower panel of Fig.~\ref{fig:iso2} shows that at $r=50\kpc$ the
differences between the distributions of $v_R$ and $v_{\rm t}$ are
qualitatively the same but quantitatively larger. 

The bottom row of panels in Fig.~\ref{fig:iso} shows a tangentially biased
model obtained by setting $\beta_v=-0.5$. The distribution of
$v_R$ is now more sharply peaked than that of $v_{\rm t}$ because
negative $\beta_v$ enhances $N(v_{\rm t})$ at
$v\sim v_{\rm c}$ while enhancing $N(v_R)$ at $v\sim0$ by enhancing the number of
circular orbits relative to eccentric orbits.

\section{Oblate models}\label{sec:oblate}

The strategy used in the last section to obtain anisotropic models of
spherical systems is not feasible in the oblate case because it requires a
model of the frequency ratios $\Omega_r/\Omega_\phi$ and
$\Omega_z/\Omega_\phi$ as functions of $\vJ$, and from section
\ref{sec:anal_flat} we know that these ratios have a complex dependence on
$\vJ$ near the plane $J_\phi=0$ (cf.~Fig~5 of \citealt{WrightB}).  Moreover,
an ergodic model cannot generate a flattened gravitational potential -- by
the virial theorem, flattening requires $\ex{v_R^2}$ to exceed $\ex{v_z^2}$.
Consequently, the strategy we pursued in the previous section to obtain
anisotropic DFs by modifying an ergodic DF, makes less sense and will not be
pursued here.

\begin{figure}
\centerline{\includegraphics[width=.9\hsize]{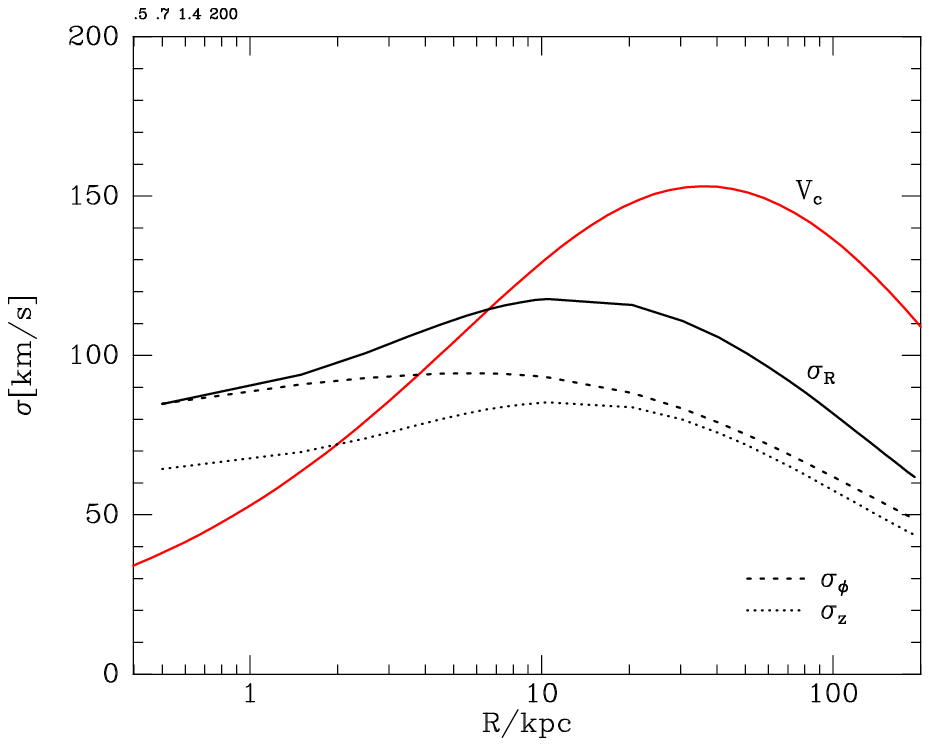}}
\caption{The principal velocity dispersions as functions of radius in a
stellar system that's trapped in an NFW potential well. The potential well is
generated by a body that has scale length
$20\kpc$ and axis ratio $c/a=0.5$. Away from the plane $J_\phi=0$, the DF is a
given by equation (\ref{eq:Posti}) with $\cJ=0.7J_r+1.4J_z+|J_\phi|$. 
In equations (\ref{eq:aux}) for the auxiliary DF and (\ref{eq:defw}) for
$w$,  the constant $\epsilon=200\kpc\kms$.}\label{fig:NFWradial}
\end{figure}

\begin{figure}
\centerline{\includegraphics[width=.8\hsize]{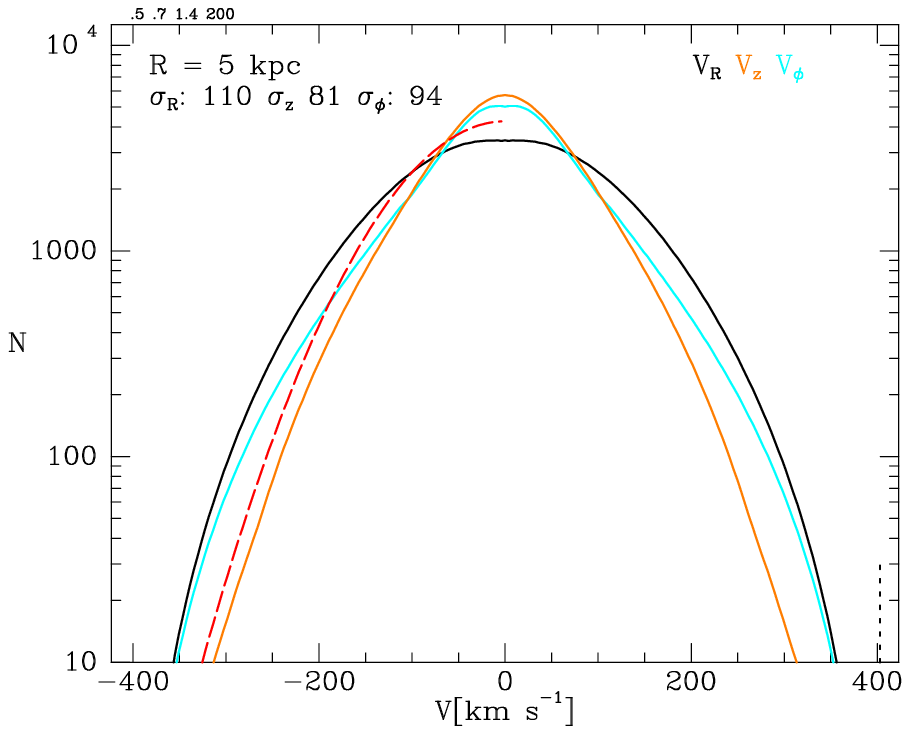}}
\centerline{\includegraphics[width=.8\hsize]{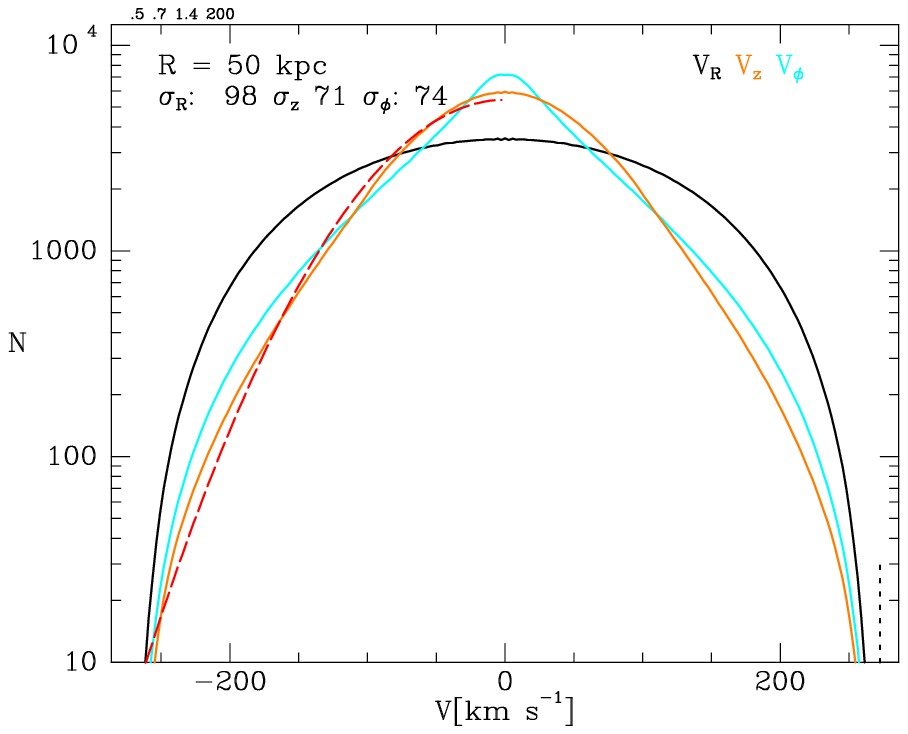}}
\caption{Velocity histograms in the equatorial plane at radii $R=5\kpc$ (upper panel) and $R=50\kpc$ (lower
panel) in the model plotted in
Fig.~\ref{fig:NFWradial}.}\label{fig:NFWvelocity} 
\end{figure}

Given a DF $f(\vJ)$, we define an auxiliary  DF $f'(\vJ)$ by
\[\label{eq:aux}
f'(\vJ)\equiv\begin{cases} 
f(J_r+\fracj12(|J_\phi|-\epsilon),J_z,\epsilon)&J_z<\Jcrit\cr
f(J_r,J_z+(|J_\phi|-\epsilon),\epsilon)& J_z>\Jcrit,
\end{cases}
\]
where $\epsilon$ is a small constant with the dimensions of action and the
relations $\Jcrit(J_r)$ and $J_{r\rm crit}(J_z)$ can be computed by
the algorithm described by \cite{WrightB} and implemented as the \agamaTwo\
function {\tt getJzcrit}.

The DF (\ref{eq:aux}) satisfies equations  (\ref{eq:condF1}) and (\ref{eq:condF2}).
Let $0\le w(\vJ)\le 1$ be a weight function such that 
\[\label{eq:wlims}
\lim_{J_\phi\to0} w(\vJ)=1 \hbox{ and }
\lim_{J_\phi\to0}\nabla_\vJ w=0,
\] 
 then the DF we use for modelling is
\[\label{eq:finalf}
f''(\vJ)=wf'(\vJ)+(1-w)f(\vJ).
\]
It's straightforward to show that
\[
\lim_{J_\phi\to0}\nabla_\vJ( f''-f')=0
\]
regardless of $f$, so $f''$, like $f'$, satisfies  equations  (\ref{eq:condF1}) and
(\ref{eq:condF2}). 


Figs.~\ref{fig:NFWradial} and \ref{fig:NFWvelocity} show profiles of a
stellar system that's trapped in the potential well of a body that has the
NFW density profile and is flattened to axis ratio $c/a=0.5$. For this figure
$f$ in equations (\ref{eq:aux}) and (\ref{eq:finalf}) took the form given by
equation (\ref{eq:Posti}) with $\cJ=0.7J_r+1.4J_z+|J_\phi|$ and
$J_0$, $\alpha,\beta,$ and $J_{\rm cut}$ as given in Table~\ref{tab:DFs}.
Fig.~\ref{fig:NFWradial} shows in black the principal velocity dispersions at
points in the equatorial plane -- note that at small radii, $\sigma_\phi$
tends to $\sigma_R$ as it should while $\sigma_z$ is everywhere smaller.
Fig.~\ref{fig:NFWvelocity} shows the velocity distributions at points in the
equatorial plane at $R=5\kpc$ (upper panel) and $R=50\kpc$ (lower panel).
Crucially, none of the histograms has a central cusp. Note that at $R=50\kpc$
the $v_z$ and $v_\phi$ histograms have quite different shapes even though
their second moments are similar.

The  weight function used to obtain Figs.~\ref{fig:NFWradial} and
\ref{fig:NFWvelocity} was
\[\label{eq:defw}
w(\vJ)={1\over1+J_\phi^2/\big[5\epsilon(5\epsilon+J_r+J_z)\big]},
\]
which satisfies equations (\ref{eq:wlims}). The model one obtains is not
sensitive to the choice of $w$.

\section{Self-consistent models}\label{sec:SelfCon}

To this point we have considered populations that are confined by given
gravitational potentials. Now we consider models in which the potential is
generated by the population itself.  

\begin{figure}
\centerline{\includegraphics[width=.9\hsize]{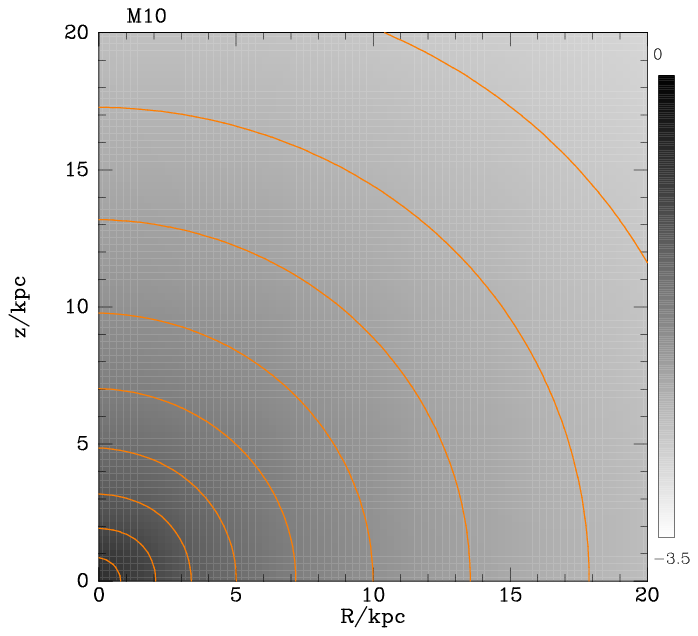}}
\centerline{\includegraphics[width=.9\hsize]{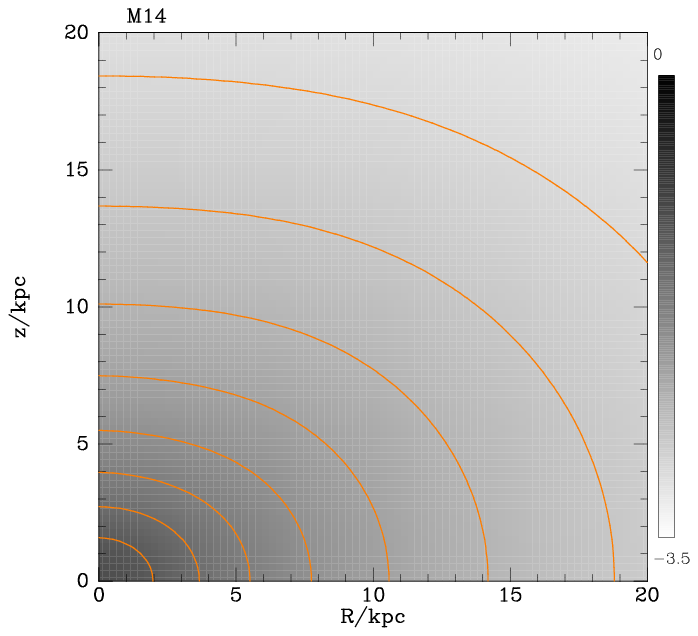}}
\caption{Two self-consistent haloes of mass $10^{12}\msun$ with DFs given by equation
(\ref{eq:Posti}) with $\alpha=1.6$, $\beta=2.7$, and $(\cJ_0,J_{\rm
cut})=(10\,000,25\,000)\kpc\kms$. For
the upper model the quantity defined by equation (\ref{eq:defcJ0}) is
$\cJ=1.4J_r+J_z+|J_\phi|$, while for the lower model it is
$\cJ=0.5J_r+J_z+|J_\phi|$.} \label{fig:M10}
\end{figure}

The upper panel of Fig.~\ref{fig:M10} shows the density $\rho(R,z)$ of a
self-consistent halo with NFW-like DF. The model's velocity distribution is
nearly isotropic: at the location of the Sun its velocity dispersions are
$(\sigma_R,\sigma_z,\sigma_\phi)=(152,150,146)\kms$ and its density
distribution is very nearly spherical. 

\begin{figure}
\centerline{\includegraphics[width=.8\hsize]{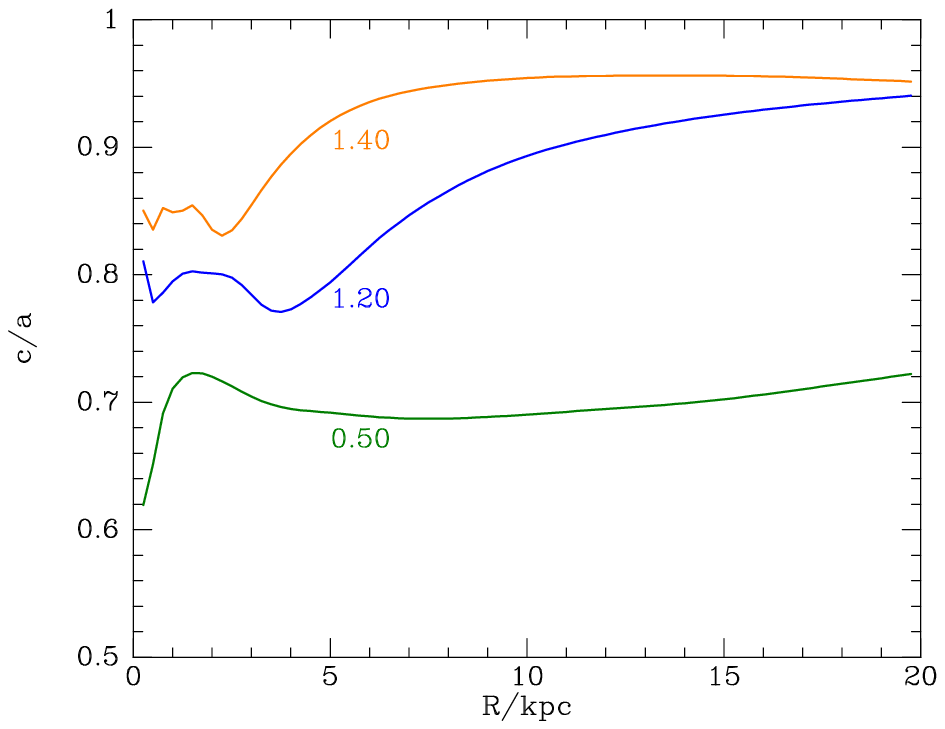}}
\caption{The axis ratio $c/a$ as a function of radius in three
self-consistent haloes with DFs based on equation (\ref{eq:Posti}).  All
three DFs  depend on $J_z$ and $J_\phi$ only through $L=J_z+|J_\phi|$. The
curves are labelled by the coefficient of $J_r$ in equation
(\ref{eq:defcJ0}).}\label{fig:ratios}
\end{figure}

The lower panel of Fig.~\ref{fig:M10} shows the result of reducing the
coefficient of $J_r$ in the definition of the quantity $\cJ$ defined by
equation (\ref{eq:defcJ0}) from $1.4$ to $0.5$.  This change creates radial
bias such that the velocity dispersions at the Sun become
$(\sigma_R,\sigma_z,\sigma_\phi)=(134,100,102)\kms$ and the model becomes
markedly oblate. Given that the DF depends on $J_z+|J_\phi|$, which, in the
case of a spherical potential, corresponds to the total angular momentum $L$,
the oblateness of this model is surprising: the natural expectation is that
the model would be spherical but radially biased rather than flattened.
Construction of a series of models in which the coefficient of $J_r$ in
equation (\ref{eq:defcJ0}) was gradually increased from $0.5$, revealed that
the model's axis ratio $c/a$ smoothly moved towards, but never quite reached
unity. Fig.~\ref{fig:ratios} shows the axis ratio as a function of radius in
the models shown in Fig.~\ref{fig:M10} and a third model in which the
coefficient of $J_r$ is $1.2$.

A natural concern is that the oblateness just reported when the DF depends on
$L$ is an artifact created by errors in the evaluation of actions by the
St\"ackel Fudge. The impact of such errors can be assessed by using the same
code to construct an isochrone sphere from the analytic DF $f[H(\vJ)]$ that's
available for the isochrone. The axis ratio $c/a$ of the resulting model
ranged from $1.02$ near the centre to $0.98$ at $20\kpc$. Thus it is very
unlikely that the oblateness evident in Fig.~\ref{fig:ratios} is an artifact.

\subsection{Instability of spherical systems}

So long as $f$ depends on $J_z$ and $J_\phi$ only in the combination $L$, a
self-consistent spherical model can be constructed for any coefficient of
$J_r$ in the definition (\ref{eq:defcJ0}) of $\cJ$. However, the following argument
shows that this model will be unstable whenever it is radially biased by virtue of
the coefficient of $J_r$ being less than $1.4$.

\cite{MayBinney1986} imagined  transferring the mass of a spherical
self-gravitating system to an external body that initially has the same
spatial distribution, so the transfer doesn't alter the distribution of the
system's (now massless) stars. When the external body is slowly made slightly
non-spherical, the stars respond adiabatically to the changed potential by
forming a non-spherical body. If this body is more aspherical than
the body that creates the gravitational field, then the original
self-consistent model is unstable.
\cite{MayBinney1986} illustrated this principle by exploring prolate
distortions of isochrone spheres with Osipkov-Merritt DFs and found that the
models remained stable until the anisotropy crossed a non-zero threshold.
Given the nature of the iterative process by which the potentials of
$f(\vJ)$ models are determined \citep{JJB14}, we can use  the argument of
\cite{MayBinney1986} to infer that the spherical analogues of the oblate
models under discussion are unstable. 

This inference suggests that radially biased spherical models are more unstable
towards flattening than towards becoming elongated: our experiments suggest
that any radial bias, no matter how small, will cause a model to be at least
a little bit oblate, whereas previous work
\citep{FridmanPolyachenkoI,PalmerPapaloizou1987,BarnesGoodmanHut,Saha1991} implies that instability against
elongation requires radial bias that exceeds a non-zero threshold.  This is
an interesting conclusion but it should be treated as tentative.
\cite{MayBinney1986} did not investigate oblate distortions. Moreover, the
models they and other authors investigated differ from the present models in two important
respects: (i) the core-envelope profiles of their isochrone models are very
unlike the double-power-law radial density profiles of the present models,
and (ii) their Osipkov-Merritt DFs create profiles of
$\beta=1-\sigma_\phi/\sigma_r$ that transition from isotropy ($\beta=0)$ inside a
characteristic radius $r_{\rm a}$ that moves inwards along their family of
DFs, to $\beta\simeq1$ at radii significantly lager that $r_{\rm a}$. The
present models have much weaker gradients in $\beta(r)$
(Fig.~\ref{fig:NFWradial}).

\section{Adiabatic deformation}\label{sec:adiabat}

The realisation that the DF of a stellar component is restricted by the
potential that it populates raises an interesting issue when the potential
is adiabatically deformed. Conventional wisdom \citep[e.g.][\S4.6.1]{GDII} is
that the function $f(\vJ)$  changes under deformation only to the extent that
orbits are resonantly trapped or chaotic. Trapping and chaos are absent in
any St\"ackel potential, so consider what happens when a
perfect ellipsoid is flattened. The flattening increases $\Jcrit$ at
any value of $J_r$ and thus changes the conditions (\ref{eq:condF1}) and
(\ref{eq:condF2}) imposed on a physical DF. Does the DF evolve so it
continues to satisfy these conditions, or does it become unphysical?

Consider orbits with vanishing $J_\phi$. Then as the potential flattens, the
natural value of $\Omega_z$ moves upwards, and when $\Jcrit$ reaches $J_z$, the
link between $\Omega_\phi$ and $\Omega_z$ breaks and $\Omega_z$ jumps up
while $\Omega_\phi$ drops to $\fracj12\Omega_r$. During the transition, at
least one of the frequency differences $\Omega_z-\Omega_\phi$ and
$\fracj12\Omega_r-\Omega_\phi$ must be small and adiabatic conservation of
the actions is not assured. Hence it is plausible that the density of stars
shifts to either maintain the conditions (\ref{eq:keyConds}) if they already
hold, or moves the DF closer to satisfying them if they do not already hold.
It would be interesting to explore this idea by numerically integrating
orbits in a slowly deforming St\"ackel potential.

\section{Conclusions}\label{sec:conclude}

Although an axisymmetric model of a stellar system can be derived from any
non-negative and normalisable function $f(\vJ)$, the model will be unphysical
unless $f$ satisfies  the conditions of equation (\ref{eq:keyConds}). These
conditions serve both to ensure that histograms of $v_\phi$ do not have cusps
at $v_\phi=0$ and to ensure that velocity distribution becomes appropriately
symmetric near the centre and near the potential's symmetry axis.

In Section \ref{sec:analytic} we saw the need for non-trivial cancellations
if $\p f/\p v_\phi$ is to vanish as $v_\phi\to0$, and by considering ergodic
DFs we showed that these cancellations require the gradients of $f$ with
respect to the actions to mirror relations between the fundamental
frequencies.  Along the way, we
established two interesting relations (\ref{eq:pJzpVphi}) and (\ref{eq:pJrpVphi})
between derivatives of actions with respect to velocities.

In Section \ref{sec:spherical} we developed a procedure for generating
ergodic DFs for spherical systems and then from these derived anisotropic DFs
that satisfy the conditions of equation (\ref{eq:keyConds}).

In Section \ref{sec:oblate} we tackled the harder problem posed
by flattened systems by defining an auxiliary DF that satisfies the
constraints near $J_\phi=0$, and then defining the DF to be a linear
combination of the original and auxiliary DFs that tends to the auxiliary DF
as $J_\phi\to0$. We showed that a component with one of these DFs that
resides in an externally generated potential can have physically acceptable
kinematics: $\sigma_z$ is always less that $\sigma_R$ while $\sigma_\phi$ is
smaller than $\sigma_R$ at large radii but tends to $\sigma_R$ as the centre
is approached. Moreover, in these models the $v_\phi$ distributions don't
have cusps at $v_\phi=0$.

We finished by constructing self-consistent models with DFs that depend on
$\cJ=aJ_r+L$, where $a$ is a free parameter and $L=J_z+|J_\phi|$. When
$a\le1.4$, the model is radially biased, the bias increasing as $a$
diminishes. To our surprise, all radially biased models proved to be oblate
-- there appears to be no space for radially biased but spherical models.
Concern that this result is an artifact produced by use of the St\"ackel
Fudge to evaluate actions was allayed by the near sphericity of the isochrone
model that the code constructed from the isochrone's known analytic
ergodic DF $f(\vJ)$. 

The oblateness of radially biased models with $f(J_r,L)$
suggests that all radially biased spherical models are unstable to
quadrupolar perturbations, but this possibility needs to be confirmed by
considering a wider range of density profiles and use of linear-response
theory to test for instability.

\cite{BinneyVasiliev2023,BinneyVasiliev2024} used the \agama\ software
package to fit data from the second and third Gaia data releases and the
17\,th data release from the APOGEE survey with models in which each of our
Galaxy's components is defined by a DF $f(\vJ)$. The DFs of the stellar and
dark haloes were chosen such that these components have nearly isotropic
velocity distributions in order to avoid the unphysical features described in
Section~\ref{sec:problem}.  The new DFs open the way to producing much more
realistic models of our Galaxy's stellar halo and to tightening constraints
on the structure of the dark halo. N-body simulations of the clustering of
dark matter strongly suggest that dark haloes are radially biased.  From this
fact and the results of Section \ref{sec:SelfCon} it follows that the dark
halo should be quite oblate. Given that the halo's contribution to the
circular-speed curve $\Vc(R)$ is already tightly constrained and flattening a
halo at fixed mass increases $\Vc$, a flatter halo has to be less massive.
Stellar streams, large numbers of which have been found in Gaia data
\citep{Ibata2024}, should be able to measure the halo's mass and oblateness
with some precision.

\section*{Acknowledgements}

I thank C.\ Nipoti and R.\ Pascale for comments on an early draft of this
paper.

\section*{Data Availability}

The computations were performed by C++ code linked to the \agamaTwo\ library
\citep{BinneyVasilievWright}, which can be downloaded from {\tt
https://github.com/binneyox/AGAMAb}.

\def\physrep{Phys.~Reps}
\bibliographystyle{mn2e} \bibliography{/u/tex/papers/mcmillan/torus/new_refs}

\appendix

\section{Limiting form of $\p J_r/\p v_{\rm t}$}\label{app:limit}

Here we infer the limit as $v_{\rm t}\to0$ of equation (\ref{eq:pJrpvphi}). 
\[
{\p J_r\over\p v_{\rm t}}\bigg|_r={v_{\rm t}\over\pi}\int_{r_{\rm p}}^{r_{\rm a}}\d r'\,
{1-r^2/r^{\prime2}\over\sqrt{2(E-\Phi)-L^2/r^{\prime2}}}.
\]
Since the second term in the numerator dominates as $r_{\rm p}$ tends to
zero, we retain only this term. We extract a factor $L$ from the
denominator and $r^2$ from the surviving numerator and change integration
variable from $r'$ to $u\equiv1/r'$. Then we have
\[
{\p J_r\over\p v_{\rm t}}\bigg|_r\simeq-\sgn(v_{\rm t}){r\over\pi}\int_{u_{\rm a}}^{u_{\rm
p}}
{\d u\over\sqrt{a^2-u^2}},
\]
where
\[
a^2=2{E-\Phi(r')\over L^2}.
\]
We neglect the weak $u$ dependence of $a$ since the integral is dominated by
the region $u\simeq a\to\infty$ just as the original integral was dominated by the
region $r'\simeq r_{\rm p}\to0$. Then, with $x=u/a$ the integral tends to
\[
{\pi\over2}=\int_0^1{\d x\over\sqrt{1-x^2}}
\]
and equation (\ref{eq:dJrdv}) follows.

\end{document}